\documentstyle[psfig,11pt,newpasp,twoside]{article}
\index{summary}
\index{instructions}
\index{template}

\def\edcomment#1{\iffalse\marginpar{\raggedright\sl#1\/}\else\relax\fi}
\reversemarginpar

\begin{document}
\title{Galactic SNR candidates in the ROSAT all-sky survey}
 \author{Daniel Schaudel, Werner Becker, Wolfgang Voges, Bernd Aschenbach}
\affil{MPE, Giessenbachstrasse 1, 85740 Garching, Germany}
\author{Wolfgang Reich}
\affil{MPIfR, auf dem H\"ugel 69, 43121 Bonn, Germany}
\author{Martin Weisskopf}
\affil{MSFC, Space Sciences SD50, Huntsville, Alabama 35812, USA}
\begin{abstract}
{\sl
Identified radio supernova remnants (SNRs) in the Galaxy comprise an incomplete sample of the SNR
population due to various selection effects. ROSAT performed the first all-sky survey with an imaging
X-ray telescope, and thus provides another window for finding SNRs and compact objects that may reside
within them. Performing a search for extended X-ray sources in the ROSAT all-sky survey database about
350 objects were identified as SNR candidates in recent years (Busser 1998). Continuing this systematic
search, we have reanalysed the ROSAT all-sky survey (RASS) data of these candidates and correlated the
results with radio surveys like NVSS, ATNF, Molonglo and Effelsberg. A further correlation with SIMBAD
and NED was performed for subsequent identification purposes. About 50 of the 350 candidates turned out to be
likely galaxies or clusters of galaxies. We found 14 RASS sources which are very promising SNR candidates
and are currently subject of further follow-up studies. We will provide the details of the identification
campaign and present first results.
}
\end{abstract}

\section{Introduction}
 Supernovae (SNe) are rare events, believed to occur at intervals of $\sim$30-50~years in the
 Galaxy (van den Berg \& Tamman 1991; Tamman et al. 1994). However, in the past 2000 years, only 7
 Galactic SNe have been observed~--- SN 185 (RCW86), SN 386 (G11.2-0.3), SN 1006, SN 1181 (3C58),
 Crab SN, Tycho SN, and Kepler SN. Most Galactic SNe appear unobserved owing to visible-band extinction
 by interstellar dust. When observational techniques in the radio band became available and the SNR
 Cas~A was found to be the brightest radio source in the sky, several surveys and directed searches for SNR
 were performed at decimeter-wavelengths. Of course, radio observations are unhampered by interstellar dust.
 Currrently, about 250 SNRs have been identified in the radio band (Green 2000 and references therein).
 Although this catalog represents the result of more than 40 years of intensive search with the largest radio
 telescopes, it is incomplete and strongly biased by mainly two selection effects:
 (i) the surface brightness
 of the remnant must be above the sensitivity limit of the observations, and (ii) the angular size of the remnant
 must be at least several times the resolution of the observations. These effects mean that not only are old faint
 remnants missing in the current catalogues, but there is also a deficit of young but distant SNRs (Green 1991).
 This is only a small fraction of SNRs of the total number expected according to the galactic SNe rate and the
 life-time of their remnants.
 \newline
 The RASS was performed from 1990 June to 1991 February and is the first
 (and only) all-sky survey done with an imaging X-ray telescope and thus provides another window that can be
 exploited to find SNRs as well as the compact objects that may reside within them (Aschenbach 1996). ROSAT was sensitive
 at soft X-ray energies (0.1-2.4~keV), had an angular resolution in survey mode of $\sim$$96^{''}$ and
 a limiting survey sensitivity of $f_x\sim 3 \times 10^{-13}$ erg~cm$^{-2}$~s$^{-1}$ (Voges et al. 1999).
 The exposure time varied between about 400 s and 40,000 s in the ecliptic plane and poles respectively.

\section{Analysis and Results}
Performing a search for extended ($\geq$ 5 $\arcmin$) and unidentified X-ray sources in the RASS database
at $\left|b \right|\leq $ 15
$\deg$, about 350 objects were identified as potential SNR candidates (Busser 1998). Studying their X-ray morphology
and correlating the results with radio catalogues and surveys like NVSS, ATNF, Molonglo and Effelsberg and the optical DSS as well
as with SIMBAD and NED, allowed to discriminate between extragalactic background objects and SNR candidates, leaving
230 targets for subsequent identification purpose. About 50 of the 350 candidates turned out to be likely galaxies or
clusters of galaxies, whereas $\sim$ 70 were found to be spurious background features. So far, 14 RASS sources
were found to be promising SNR candidates and are currently subject of further follow-up studies in the X-ray
and radio band. Due to the relatively low photon statistics and spatial resolution of the RASS data, it is not possible
to perform a detailed analysis which allows to identify the nature of the objects. Below we briefly mention three targets
which turn out to be likely SNRs according to their X-ray and radio morphology.
\begin{itemize}
\item G296.7-0.9 is located in a HII region (Russeil 1996) which is overlapping partly the area of radio emission.
The target has an extent of about 15\arcmin$\times$10\arcmin ~in X-rays and at radio wavelength.
A pointed ROSAT HRI observation which has a 18 times higher angular resolution compared to the survey observation shows
an incomplete shell in the east, supporting strongly the SNR interpretation (Fig.~1a/b).
\item G308.3-1.4 is listed as a possible SNR in the MOST supernova remnant catalogue based on the
composite  radio morphology and the ratio of 60 $\mu$m to 843 MHz flux densities (Whiteoak 1992). The RASS PSPC data 
support this identification showing a hard X-ray source with a diameter of about 10\arcmin ~ partly overlapping with 
the radio arcs (Fig.~1c).
\item G38.7-1.4 shows X-ray emission in the RASS with an extent of about
12\arcmin$\times$8\arcmin, surrounded
by an incomplete radio shell in the Effelsberg galactic plane 11~cm-survey (Reich et al 1990).
A follow-up observation with the Effelsberg 100-m telescope at 6~cm wavelength
was recently performed, which shows the shell to be significantly polarized.
This result clearly favours a SNR interpretation of G38.7-1.4 (Fig.~1d).
\end{itemize}
\begin{center}
\begin{figure}[h]
\psfig{figure=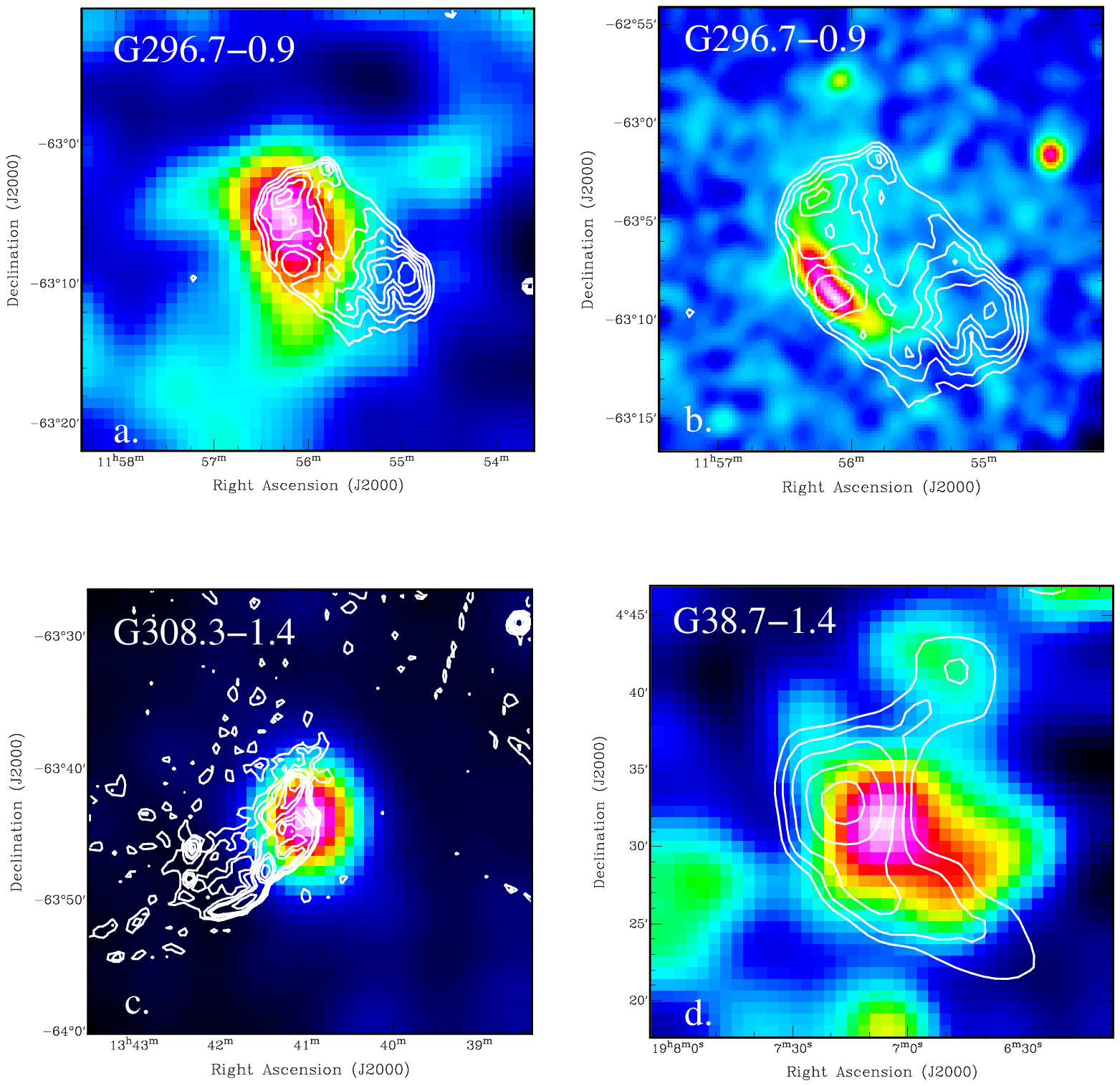,height=12.9cm,width=12.9cm,angle=0}
\hspace{1.0cm}
\vspace{0.5cm}
\centerline{\renewcommand{\baselinestretch}{1.05}\parbox{13.5cm}{\small {\bf Fig.~1}{\bf a.} Greyscaled RASS image of
G296.7-0.9 with superimposed radio contours
the MOST data at 843~MHz {\bf b.} G296.7-0.9 as seen in the ROSAT HRI shown greyscaled. The higher angular resolution 
resolves the
extended X-ray source into a faint arc structure located in the east {\bf c.} RASS image of
G308.3-1.4 overlaid with MOST data which show an incomplete radio shell {\bf d.} RASS image of G38.7-1.4 overlaid
with contour lines from the 11~cm Effelsberg survey }}
\end{figure}
\end{center}
\vspace{-1.0cm}
\section{Summary and Prospects}
About 70 galactic SNRs are listed in Green's SNR Catalogue (Green 2000) being detected in X-rays.
Making use of the spectro-imaging capability provided by ROSAT in its all-sky survey, we reanalysed a sample of about
350 extended X-ray sources, which were proposed by (Busser 1998) as SNR candidates. Performing a more dedicated imaging
analysis and making use of recently updated databases like SIMBAD and the NED  we reduced the number of SNR candidates
 to be about 230. 14 of these targets turn out to be promising SNR candidates based on available data
and are subject of follow-up observations for clarification.
Two candidates were recently observed
with the Chandra observatory and another six candidates with the Effelsberg 100-m telescope. The analysis of these
data is currently in progress. We will continue this campaign providing additional observations in X-rays with XMM-Newton
and Chandra and in radio with Effelsberg and ATNF telescopes. Optical observations will complete our multi-wavelength studies.

\end{document}